# A pulsed lock-in method for ensemble nitrogen-vacancy center magnetometry

Jixing Zhang,[1] Heng Yuan,[1,2,3,4*] Tianzheng Liu,[1] Lixia Xu,[1] Guodong Bian,[1] Pengcheng Fan,[1] and Mingxin Li.[1]

[1]School of Instrumentation and Optoelectronic Engineering, Beihang University, Beijing 100191, China;
[2]Research Institute of Frontier Science, Beihang University, Beijing 100191, China;
[3]Beijing Advanced Innovation Center for Big Data-Based Precision Medicine, Beihang University, Beijing 100191, China;
[4]Beijing Academy of Quantum Information Sciences, Beijing 100193, China;
[5]Zhejiang Lab, Hangzhou, 310000, China;
*hengyuan@buaa.edu.cn,

**Abstract**

This article proposes a scheme for nitrogen-vacancy (NV) center magnetometry that combines the advantages of lock-in detection and pulse-type scheme. The optimal conditions, optimal sensitivity, and noise-suppression capability of the proposed method are compared with those of the conventional methods from both theoretical and simulation points of view. Through experimental measurements, a four-time improvement in sensitivity and 60-times improvement in minimum resolvable magnetic field (MRMF) was obtained. By using a confocal experiment setup, proposed scheme achieves a sensitivity of 3 nT·Hz$^{-1/2}$ and a MRMF of 100 pT.

Diamond nitrogen-vacancy (NV) centers have been a hot topic in quantum sensing research in recent years. Magnetometers based on NV centers have several advantages over conventional magnetometers, including extreme environmental adaptation[1] and high sensitivity limit[2]. However, the actual sensitivity of NV magnetometers is greatly affected by technical noise and far from the sensitivity limit, especially at low frequency. For this reason, the modulation continuous-wave (CW) method combined with the lock-in (LI) detection scheme, which has great ability to surpass the low-frequency technical noise, has become the mainstream scheme in practical NV magnetometer and optical-pumped magnetometer[3,4,5]. Among the different types of NV magnetometers without flux concentrator investigated, the highest sensitivity of 15 pT·Hz$^{-1/2}$ has been achieved using this scheme[6]. Several >100 nT-level compact NV magnetometers[7,8] as well as fiber-combined probes[9] have been proposed. Being a versatile technic, this continuous-wave lock-in (CWLI) detection scheme has been combined with numerous different methods, including the closed-loop control approach, which achieves a robust high-dynamic range[10], simultaneous broadband vector magnetometry with multichannel LI detection[11], cavity-enhanced absorption detection[12,13,14,15], and microwave-free magnetometry with magnetic modulation[16]. For single NV center systems, a LI detection liked different meter has been proposed for a photon-count scenario[17]. The principle of LI detection has been further expended to quantum LI amplifiers[18] and classic clock-combined magnetic spectroscopy[19]. Aside from the CW method, the pulsed methods, such as pulsed optically-detected magnetic resonance (ODMR) and Ramsey sequences[20], exhibit higher contrast and theorical sensitivity owing to the fact that microwave (MW) manipulations are independent from laser polarization and readout[2]. However, the LI scheme has not been applied to the pulsed methods.

In this manuscript, a pulsed NV center magnetometry method is proposed combined with LI detection (abbreviated as PULI scheme) to suppress technical noise and improve sensitivity. The basic idea behind this scheme is that of setting the pulse duration to being negligible compared with the MW modulation period. Thus, in each pulse sequence, the MW frequency can be considered unchanged, and the LI detection can be applied. In this work, a mathematic model for the LI amplifier and theoretical sensitivity is proposed. The conventional pulsed and CW methods without LI detection are here abbreviated as PUUL and CWUL, respectively. Combined with the master equation, the optimal conditions, including MW power, MW modulation deviation, and corresponding optimal sensitivity, were obtained as a function of the dephasing time for the conventional CWUL, CWLI, PUUL, and PULI schemes. In addition, the noise-suppression ability was simulated for all schemes. Furthermore, the experimental system for the PULI scheme was set up, and the corresponding experimental verifications were carried on. The DC sensitivity of the PULI scheme was found to be 3 nT·Hz$^{-1/2}$, which represents a 4-fold improvement compared with the conventional scheme in same experiment setup.

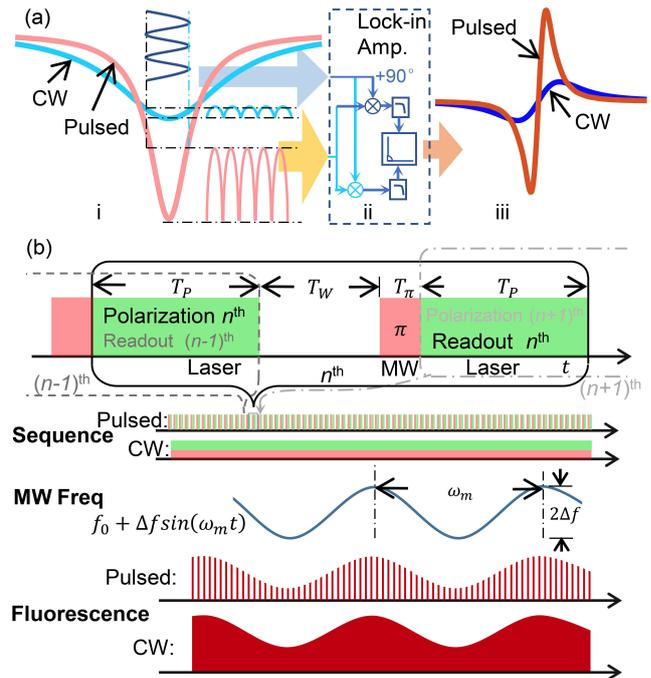

**Fig. 1** (a) Diagram illustrating the principle of LI detection for the CW and PU schemes. **i** The light red curve represents the PUOMDR spectrum, while the light blue curve represents the CWODMR spectrum. **ii** Internal structure of the LI amplifier. The first step consists of DC isolation for both the signal and reference (not shown in the figure); the reference is split into two and the phase of one of these is shifted by 90°. Both references are then multiplied by the signal and pass through a low-pass filter, resulting in $x$ and $y$. The LI amplifier output phase can then be calculated as $\theta = \arctan(x/y)$ and $R = \sqrt{x^2 + y^2}$. **iii** CWLI and PULI signals resulting from the output of the LI amplifier. The blue and red curves represent the CWLI and PULI schemes, respectively. (b) PULI sequence. In the black frame, a single measurement sequence is illustrated with duration $(T_P + T_W + T_\pi)$, where $T_P$ is the polarization pulse, $T_W$ is the waiting pulse, $T_\pi$ is the MW pulse, and $T_P$ is the readout pulse. $T_P$ is reused in the following measurement sequence, as the polarization reduces the total duration and improves the sample rate $s_{rPU}$ of the PULI scheme. The PULI fluorescence is modulated via the MW frequency modulation (the fluorescence change is exaggerated in the figure). The CWLI scheme is shown for comparison.

Benefitting from the excellent long-term stability of the PULI scheme, a minimum detectable magnetic field of 100 pT was achieved for a 1000 s measurement.

The principle of magnetic measurements using NV centers is based on the fact that the NV center energy sublevels of the electric spins in the ground state shift with magnetic field due to the Zeeman effect. Therefore, the strength of the magnetic field can be derived as $\gamma_e B + Z_{gs}$, i.e., from the shift of the ODMR spectrum due to the optical spin polarization and optical spin readout of the NV center, as shown in Fig. 1(a). For the CWUL scheme, whose sequence is shown in Fig. 1(b), $\mathcal{F}$ is set as the detected fluorescence photon number per unit time. For the PUUL scheme, $I$ is set as the total detected fluorescence photon number in a single detection pulse. At the position where the ODMR slope is maximum, i.e., $k_{CW} = \frac{d\mathcal{F}(B)}{dB}$ and $k_{PU} = \frac{dI(B)}{dB}$, ODMR-based magnetometry measurements without LI detection have the optimal sensitivity, namely

$$\eta_{PUUL} = \frac{\sqrt{T_I + T_2^* + T_R}}{k_{PU}\sqrt{I_{avg}}}$$
$$\eta_{CWUL} = \frac{1}{k_{CW}\sqrt{\mathcal{F}}}. \quad (1)$$

Based on this principle, the sensitivity of the LI detection scheme can be established. The diagram of the LI detection scheme is shown in Fig. 1(a), and can be described according to the following steps. (i) The MW frequency modulation $f_0 + \Delta f \sin(\omega_m t)$ leads to a fluorescence modulation. Here, $f_0$ is the MW frequency, $\Delta f$ is the modulation depth, and $\omega_m$ is the modulation frequency. (ii) The fluorescence and modulation signals are input into the LI amplifier as signal and reference, respectively. (iii) For the CWLI scheme, the output of the lock-in amplifier is[21]

$$S_{cw}(B) = \frac{\omega_m}{2\pi} \int_{-\frac{\pi}{\omega_m}}^{\frac{\pi}{\omega_m}} \mathcal{F}\left(f_0 + \Delta f \sin(\omega_m t) - (\gamma_e B + Z_{gs})\right) \sin(\omega_m t) dt$$
$$\approx \frac{\omega_m}{2\pi} \int_{-\frac{\pi}{\omega_m}}^{\frac{\pi}{\omega_m}} \mathcal{F}'_{f_0}(B) \Delta f \sin^2(\omega_m t) dt = \frac{\mathcal{F}'_{f_0}(B)\Delta f}{2}, \quad (2)$$

which corresponds to the ODMR derivative. Here, integration is utilized to represent the low-pass filter, and the optimal modulation depth is

$$\Delta f = \sqrt{\frac{-8\mathcal{F}_{f_0}^{(3)}}{\mathcal{F}_{f_0}^{(5)}}}. \quad (3)$$

The corresponding optimal theorical sensitivity is

$$\eta_{CWLI} = \frac{\sqrt{2\mathcal{F}}}{\gamma_e \sqrt{\frac{-2\mathcal{F}_{f_0}^{(3)}}{\mathcal{F}_{f_0}^{(5)}}}\left(\mathcal{F}_{f_0}^{''}(B) - \mathcal{F}_{f_0}^{(4)}\frac{\mathcal{F}_{f_0}^{(3)}}{\mathcal{F}_{f_0}^{(5)}}\right)}. \quad (4)$$

For the PULI scheme, using a similar derivation, the optimal theorical sensitivity is (see Supplement for derivations of Eq. (3-5))

$$\eta_{PULI} = \frac{\sqrt{T_I + T_2^* + T_R}\sqrt{2I}}{\gamma_e \sqrt{\frac{-2I_{f_0}^{(3)}}{I_{f_0}^{(5)}}}\left(I_{f_0}^{''}(B) - I_{f_0}^{(4)}\frac{I_{f_0}^{(3)}}{I_{f_0}^{(5)}}\right)}. \quad (5)$$

In the theorical analysis, the conditions of the optimal modulation frequency and low-pass filter parameters are ideal. These conditions will be discussed in the following simulations and measurements. As shown in Fig. 2(a), the master equation model for analyzing the dynamics of the NV center states under MW and laser conditions is written as

$$\begin{aligned}
\dot{\rho}_{11} &= -\Gamma\rho_{11} + k_{31}\rho_{33} + k_{51}\rho_{55} + i\Omega(\rho_{12} - \rho_{21})/2 - (\rho_{11} - \rho_{22})/T_1 \\
\dot{\rho}_{22} &= -\Gamma\rho_{22} + k_{42}\rho_{44} + k_{52}\rho_{55} + i\Omega(\rho_{21} - \rho_{12})/2 - (\rho_{22} - \rho_{11})/T_1 \\
\dot{\rho}_{33} &= \Gamma\rho_{11} - (k_{35} + k_{31})\rho_{33} \\
\dot{\rho}_{44} &= \Gamma\rho_{22} - (k_{45} + k_{42})\rho_{44} \\
\dot{\rho}_{55} &= k_{35}\rho_{33} + k_{45}\rho_{44} - (k_{51} + k_{52})\rho_{55} \\
\dot{\rho}_{12} &= (-1/T_2^* + i\delta)\rho_{12} + i\Omega(\rho_{11} - \rho_{22})/2 \\
\dot{\rho}_{21} &= (-1/T_2^* + i\delta)\rho_{21} + i\Omega(\rho_{22} - \rho_{11})/2
\end{aligned} \quad (6)$$

$\Gamma$ represents optical pumping rate, $\rho_{ij}$ represents the $ij$th terms of density matrix, $k_{ij}$ represents the relaxation rates from $i$ state to $j$ state, $T_2^*$ is the dephasing time, $T_1$ is the lattice relaxation time, $\Omega$ is Rabi frequency, and $\delta = f_0 - (\gamma_e B + Z_{gs})$ is detuning frequency. The fluorescence dynamic model and corresponding parameters are taken from Ref. [20]. Since the $T_2^*$ dephasing time plays an important role in the optimal conditions and sensitivity, schemes with different $T_2^*$ were simulated. Firstly, the duration of each part of the PULI sequence was confirmed. The waiting time was set as $T_W \geq 400$ ns in order to ensure a full relaxation for the excited and singlet states. The polarization time was set as $T_P \geq 2$ μs to ensure that the spin substate repolarizes to $m_s = 0$ as well as to obtain a sufficiently duration for

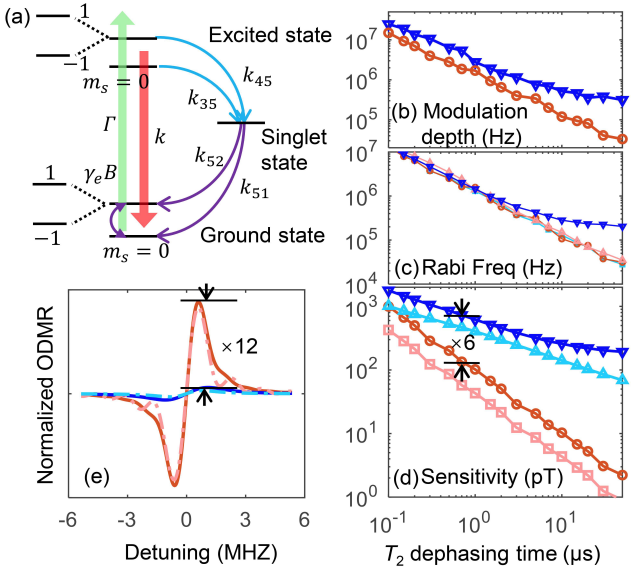

**Fig. 2** (a) Diagram of the NV optical dynamic model. (b) Relationship between the optimal modulation depth and $T_2^*$. (c) Relationship between the optimal MW power (Rabi frequency) and $T_2^*$. (d) Relationship between the optimal sensitivity and $T_2^*$. (e) ODMR signal for the four schemes at $T_2^*$ = 800 ns. The dark blue curve and "▼" symbol represent the CWLI scheme, the light blue curve and "▲" symbol represent the CWUL scheme, the dark red curve and "o" symbol represent the CWLI scheme, and the light red curve and "□" symbol represent the PUUL scheme.

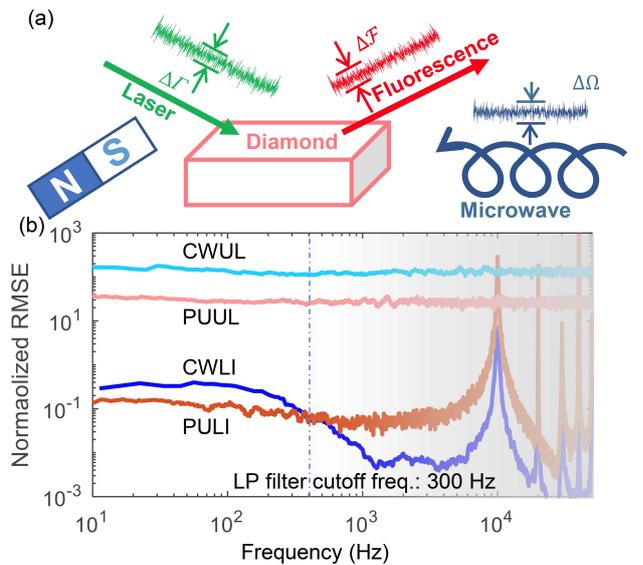

**Fig. 3** (a) Diagram of the different noise sources. The green features and $\Delta\Gamma$ represent the laser noise and its RMSE; the red features and $\Delta\mathcal{F}$ represent the detection noise and its RMSE; the dark purple features and $\Delta\Omega$ represent the MW power noise and its RMSE. (b) Normalized RMSE of the magnetic field noise for the different schemes under the same laser noise.

spin readout. The MW time is a Rabi $\pi$ pulse and varies from 10 ns to 1.5 μs depending on the MW power. Considering that a smaller total time is beneficial for the sensitivity, the total duration was set to 4 μs, and the corresponding sample rate $s_{rPU}$ was set to 250 kHz. The modulation frequency $\omega_m$ was set 10 kHz, a value at which the response of the NV fluorescence does not decline, as experimentally verified (see supplementary material). The low-pass filter cutoff frequency of the LI amplifier was set to 300 Hz. Considering the similar sensitivity performance of the pulsed-ODMR and Ramsey scheme[22], it is not necessary to investigate the Ramsey scheme in a particular way.

For the PUUL and CWUL schemes, the optimal MW power is represented by $\Omega$ and corresponding $\delta$, which are coupled with each other. Similarly, for the PULI and CWLI schemes, the optimal MW power $\Omega$ and optimal modulation depth $\Delta f$ are coupled. Thus, it is necessary to simulate the two parameters together and find the optimal pair for $(\Omega, \delta)$ or $(\Omega, \Delta f)$. (see supplementary material for more details). The optimal conditions and corresponding optimal sensitivity are shown in Fig. 2(b-d). It can be seen that the PULI scheme has a smaller $\Delta f$, since the PU scheme has a smaller ODMR width due to the optically-induced broadening. The optimal MW power is similar among the four schemes, except for the CWLI scheme. The sensitivities of the LI schemes are worse than those of the direct measurements, which is due to the information loss of the LI detection. The sensitivities of the CW schemes are worse than that of the pulsed schemes, and as the $T_2^*$ increases, the difference becomes larger. This difference can be attributed to the optical broadening and optically-induced decrease in contrast.

The ODMR derivative for the four schemes at $T_2^* = 0.8$ μs was calculated, as shown in Fig. 2(e), considering that the $T_2^*$ of the investigated diamond sample is ~0.8 μs. Compared with the CW contrast, it can be seen that the pulsed contrast is ~12 times higher. However, the sensitivity can only be increased by ~6 times. This is also consistent with the predictions of the sensitivity equation. In the PU schemes, only the fluorescence of the pulsed part is used for the spin readout, while the whole fluorescence signal carries magnetic field

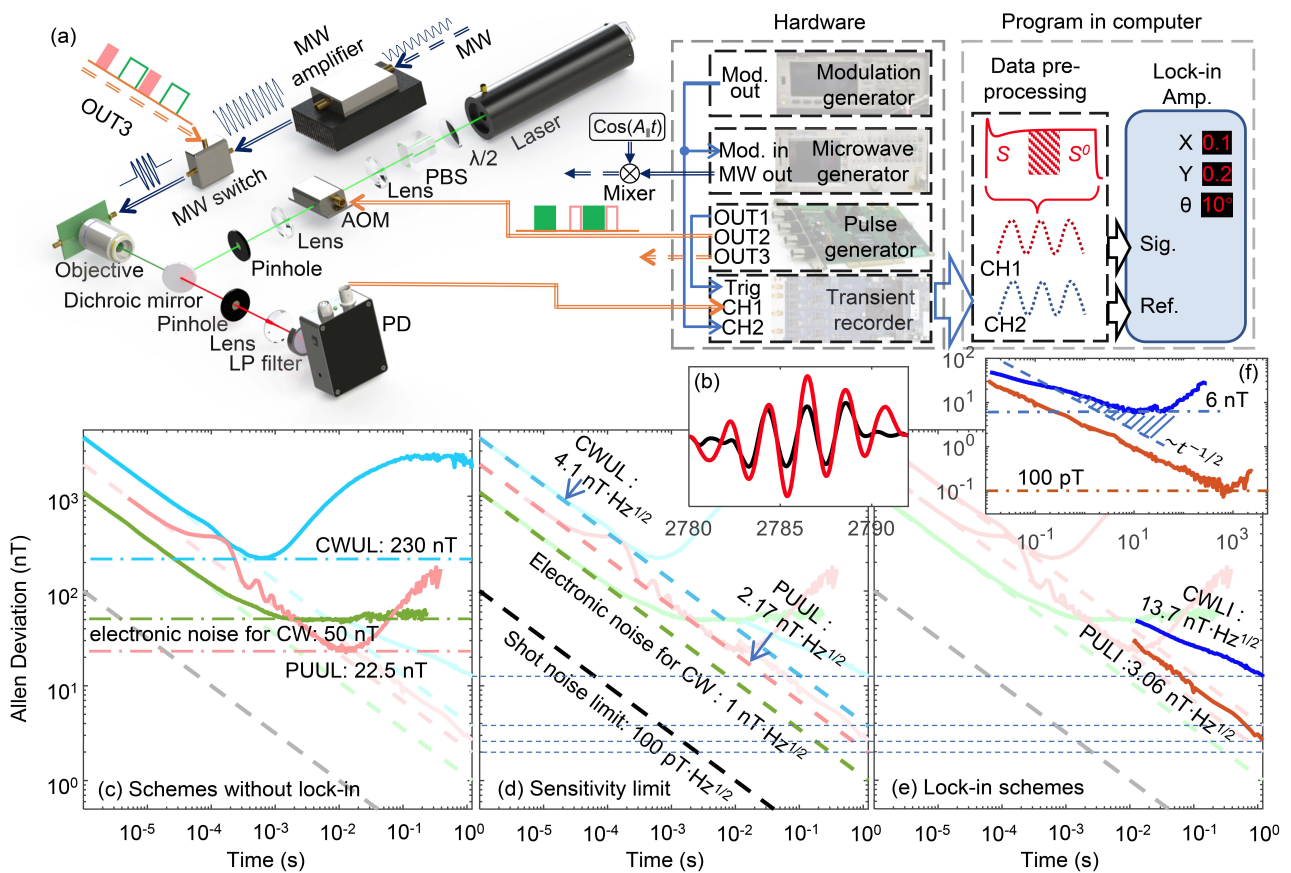

**Fig. 4** (a) Schematic diagram illustrating the experiment setup. The left, middle, and right parts are the confocal optical path, main hardware, and diagram of algorithm implementation, respectively. For the optical design part, the dichroic mirror separates the fluorescence from the laser. The polarizing beam splitter (PBS) and λ/2 waveplate are used to adjust the laser power. The acousto-optic modulator (AOM) and pinhole are used to generate the laser pulses, and the two lenses next to the AOM are used to focus the laser spot to improve the AOM diffraction efficiency. The diamond and MW antenna are blocked by the objective lens. The 800 nm long-pass filter is used to further cutoff the laser. The lens and pinhole before the photodetector (PD) are used to improve the fluorescence beam quality. For the hardware part, all signal flows are indicated by the arrows. The MWs outputted from the MW generator are frequency-modulated, relying on the modulation signal. Subsequently, in order to employ three hyperfine levels of nuclear $^{14}$N, the MW signal is split into three frequencies by mixing with a radiofrequency signal, whose frequency is given by the hyperfine separation $A_1 = 2.16$ MHz. The OUT1 for the pulse generator is connected to the trigger of the transient recorder to start the data recording process. OUT2 and OUT3 are connected to the AOM and MW switch for the laser and MW pulses, respectively. For the programming part, the fluorescence signal is firstly pre-processed. The first 500 data points (referring to the first 1 μs) of the laser pulse serve as the spin readout signal S and last 500 datapoints (referring to the last 1 μs) of laser pulse serve as the reference S0. Based on the spin readout methods from Refs. [23][20], the spin readout results for each measurement sequence are obtained and input into the LI amplifier as signal. The reference for the LI amplifier is taken from the modulation signal obtained using the same sampling rate as the spin readout results. (b) The LI amplifier signals for the PULI scheme without mixing cos($A_1t$) (black curve) and with mixing cos($A_1t$) (red curve) show a 2-fold improvement in the scale factor for the triple-peak simultaneous manipulation with the mixer. (c-f) Allen deviations of the PULI and conventional schemes. (c) Highlighting the Allen deviation for the CWUL and PUUL schemes, and the PD electronic noise for the CWUL scheme, as well as their minimum detectable magnetic field. (d) Highlighting the sensitivity at high frequency for the CWUL, PUUL, electronic noise, and shot noise limit. (e) Highlighting the Allen deviation for the CWLI and PULI schemes and their corresponding sensitivity. (f) Allen deviation reaching 5000 s for the CWLI and PULI schemes as well as their minimum detectable magnetic field.

information for the CW schemes; thus, the sensitivity is not improved as significantly as the ODMR contrast is.

Although the LI schemes have lower sensitivity, their advantage lies in the high technical noise suppression ability, which was confirmed via simulations, as shown in Fig. 3. The dominant technical noise sources are the laser noise, detection noise, and MW power noise, as shown in Fig. 3(a). Taking the laser noise as an example, simulations were carried out by replacing the fixed laser power $\Gamma$ with the white noise containing laser power $\Gamma(t)$ and by importing $\Gamma(t)$ in the dynamic model of Eq. (6) for the different schemes. The comparison of the root mean squared error (RMSE) for the magnetic signal of the different schemes can be used to express the comparison of the technical noise suppression ability. In this simulation, the PU schemes utilize a differential approach, as shown in Fig. 4(a). The results obtained for the suppression ability, shown in Fig. 3(b), demonstrate that the LI schemes perform ~100 times better than the schemes lacking LI detection. This is due to the equivalent band-pass filter. The PU schemes are 10 times better than the CW schemes owing to their differential approach. The results obtained for the other noise sources are similar to the those here discussed for the laser noise, so they are not presented. To conclude, the simulation results show that the PULI scheme can nearly achieve the highest theoretical sensitivity and best technical noise suppression ability; thus, the PULI scheme can be considered as optimal compared with the conventional schemes.

The main experiment setup is shown in Fig. 4(a), and the equipment used is described in the following. The transient recorder was a spectrum M4i.4450-x8 with sample rate of 500 MHz, the microwave generator was an Agilent MXG N5181B, the modulation generator was an Agilent 33500B, and the pulse generator was a SpinCore PBESR-PRO-500. The diamond sample was fabricated via chemical vapor deposition (CVD) using 50 ppm N impurity under $1 \times 10^{18}$ e/cm$^2$ and 10 MeV electron irradiation. Subsequently, the sample was annealed for 2 h at 800°C. The NV density was ~3 ppm. A bias magnetic field of 30 G was applied along the (111) direction. This field was generated via either a permanent magnet or a three-dimensional Helmholtz coil, which are not shown in Fig. 4(a).

The optimal parameters, including modulation frequency, waiting time, and MW power, were experimentally demonstrated. The experimental results were found to be in good agreement with the simulation results (see supplementary material for details). An Allen deviation was utilized to analyze the performance of the PULI and conventional schemes. The bottom of the Allen deviation represents the minimum detectable magnetic field (MDMF), while the intersection of the Allen deviation curve with the $x = 1\ s$ line represents the sensitivity at 1 Hz. In the high-frequency range, the sensitivity is represented by the intersection of the extrapolated -1/2 slope line of the Allen deviation curve with the $x = 1\ s$ line, as shown in Fig. 4(d). The sensitivity at high frequency for the CWUL and PUUL schemes are 4.1 and 2.2 nT·Hz$^{1/2}$, respectively, while the corresponding MDMF values are 230 and 22.5 nT. The sensitivity at 1 Hz for the CWUL and PUUL schemes are considerably worse. The sensitivity at 1Hz for CWLI and PULI are 13.7 and 3.06 nT·Hz$^{1/2}$, respectively. The Allen deviation for the CWLI scheme does not decrease with -1/2 slope, which could be attributed to the laser noise, and its MDMF value is 6 nT. Owing to the superior low-frequency technical noise suppression ability, the MDMF of the PULI scheme can reach 100 pT at 1000 s.

In summary, a novel scheme for NV center magnetometry, which combines LI detection and pulsed approach, was proposed in this work. The proposed scheme exhibits both the high technical noise suppression ability typical of LI detection and the high sensitivity typical of the pulsed method. The mathematical model for this scheme was implemented. Based on this model, the scheme was fully analyzed and simulated. The experimental results are in good agreement with the theorical results. The proposed method exhibits a 4-fold improvement in sensitivity and 60-fold improvement in the minimum detectable magnetic field compared with the conventional CWLI detection scheme. Indeed, a sensitivity of 3 nT·Hz$^{1/2}$ at 100-0.001 Hz and minimum detectable magnetic field of 100 pT were demonstrated. Since the experimental investigations focused on the verification of the proposed method and its comparison with the conventional schemes, a confocal optical path was used, causing a smaller sample volume and lower number of NV centers to participate in the measurement. Thus, when applied to NV magnetometer scheme with high participant sample volume[6], the proposed scheme could lead to a sub 10 pT·Hz$^{1/2}$ sensitivity and sub pT minimum detectable magnetic field.